\documentclass[twocolumn, prl, superscriptaddress, notitlepage]{revtex4-1}
\usepackage{graphicx}
\usepackage{physics}
\usepackage{color}
\usepackage{amsmath}
\usepackage{amsfonts}
\usepackage[normalem]{ulem}
\usepackage{xspace}
\usepackage[dvipsnames]{xcolor}
\usepackage{dcolumn}
\usepackage[breaklinks,colorlinks,linkcolor=blue,citecolor=blue,urlcolor=blue]{hyperref}

\begin{document}
\title{Synthetic tensor gauge fields}

\author{Shaoliang Zhang}
\email{shaoliang@hust.edu.cn}
\thanks{They contribute equally to this work.}
\affiliation{School of Physics, Institute for Quantum Science and Engineering, Huazhong University of Science and Technology, Wuhan 430074, China}

\author{Chenwei Lv}
\email{lvc@purdue.edu}
\thanks{They contribute equally to this work.}

\affiliation{Department of Physics and Astronomy, Purdue University, West Lafayette, IN, 47907, USA}

\author{Qi Zhou}
\email{zhou753@purdue.edu}
\affiliation{Department of Physics and Astronomy, Purdue University, West Lafayette, IN, 47907, USA}
\affiliation{Purdue Quantum Science and Engineering Institute, Purdue University, West Lafayette, IN 47907, USA}
\date{\today}

\begin{abstract}
Synthetic gauge fields have provided physicists with a unique tool to explore a wide range of fundamentally important phenomena. 
However, most experiments have been focusing on synthetic vector gauge fields. The very rich physics brought by coupling tensor gauge fields to fracton phase of matter remain unexplored in laboratories.  
Here, we propose schemes to realize synthetic tensor gauge fields that address dipoles instead of single-particles.  
A lattice tilted by a strong linear potential and a weak quadratic potential yields a rank-2 electric field for a lineon formed by a particle-hole pair.  
Such a rank-2 electric field leads to a new type of Bloch oscillations, which modulate the quadrupole moment and preserve the dipole moment of the system.
In higher dimensions, the interplay between interactions and vector gauge potentials imprints a phase to the ring-exchange interaction and thus generates synthetic tensor gauge fields for planons. 
Such tensor gauge fields make it possible to realize a dipolar Harper-Hofstadter model in laboratories. 
The resultant dipolar Chern insulators feature chiral edge currents of dipoles in the absence of net charge currents. 
\end{abstract}
\maketitle 

The study of couplings between matter and gauge fields is an essential topic in physics, telling us how our universe functions at all length scales
~\cite{Lasenby1998, Thouless1982, Tsui1982, Konig2007}. 
Experimentalists have realized synthetic gauge fields using ultracold atoms, photonics and other highly controllable systems~\cite{Lin2011, Bloch2011, Bloch2013, Ketterle2013, Schweizer2019, Grg2019, Yang2020, Chen2022, Zhou2022, Hafezi2011, Khanikaev2013, Hafezi2014, Ssstrunk2015, Xiao2015, Lee2018, SynDim_1, Lewenstein2014, SynDim_2, SynDim_3}. 
Whereas these synthetic gauge fields have allowed physicists to explore fundamental problems in high-energy physics and simulate novel topological quantum matter, most experiments have been focusing on vector gauge fields. 
On the other hand, tensors could bring us with even richer physics, ranging from tensor-momentum coupling~\cite{Cui2013} to novel single-particle band structures produced by synthetic tensor monopoles~\cite{Palumbo2018, Chenmo2022, Tan2021, Zhuyan2020}.
In particular, tensor gauge fields are crucial in the study of fracton phases of matter~\cite{Chamon2005, Bravyi2011, Haah2011, Yoshida2013, Fu2015, Fu2016, Pretko2017, Pretko2017_2, Nandkishore2019, Pretko2020}, where a single fracton is immobile~\cite{Scherg2021, Senthil2022, Senthil2023, Bloch2023}, and a dipole formed by a particle-hole pair could move. 
Some dipoles only move in the direction parallel to the dipole moments and are referred to as lineons.
In contrast, planons move in directions perpendicular to the dipole moments. 

Whereas extensive theoretical studies have shown that coupling lineons and planons to tensor gauge fields leads to exotic phenomena unattainable in traditional platforms~\cite{Pretko2017_3, Pretko2018, Chen2018, Barkeshli2018}, 
a bottleneck is the absence of such tensor gauge fields in experiments. 
For instance, the kinetic energy of a lineon in a bosonic system reads, 
\begin{equation}
    \hat{K}_L=-\sum_{m}(t_{2}e^{-i A_{xx}}\hat{b}^{\dagger 2}_{{ m}}\hat{b}_{ m+1}\hat{b}_{ m-1}+h.c.),\label{Kdp}
\end{equation}
where $\hat{b}^{\dagger }_{{ m}}$ ($\hat{b}_{{ m}}$) is the creation (annihilation) operator for bosons at site ${ m}$ and $t_2$ is the strength of the correlated tunneling.  
$A_{xx}$ is a rank-2 gauge potential and couples to the second derivative of the bosonic field, since the kinetic energy $\sim (A_{xx} -a^2\partial^2_x\theta_B)^2$, where $a$ is the lattice spacing, $\theta_B$ is the phase of the bosonic field and $x$ is the coordinate of the lattice in the continuum limit. 
When $A_{xx}=0$, Eq. (\ref{Kdp}) denotes correlated tunnelings that have been realized in experiments~\cite{Yang2020, Scherg2021, Zhou2022, Bloch2023}. 
However, an intrinsic challenge is how to access a finite $A_{xx}$ and other tensor gauge fields in experiments. 
The existing schemes in the literature could only gauge single-particle tunnelings and realize terms like $e^{-i A_{x}}\hat{b}^{\dagger }_{{ m}}\hat{b}_{ m+1}+h.c.$, where $A_x$ is a rank-1 gauge field. 
Without creating a finite $A_{xx}$ and other tensor gauge fields in experiments, it is impossible to test profound theoretical predictions of how fracton phases of matter interact with tensor gauge fields.

Here, we point out schemes to realize synthetic tensor gauge fields using currently available experimental techniques and demonstrate a range of unique quantum phenomena brought by synthetic tensor gauge fields.
For instance, a rank-2 electric field creates a new type of Bloch oscillations dubbed as dipolar Bloch oscillations. 
Distinct from conventional Bloch oscillations changing the dipole moment of the system, a dipolar Bloch oscillation periodically modulates the quadrupole moment while preserving the dipole moment. 
A higher rank magnetic field allows experimentalists to access a dipolar Harper-Hofstadter model and dipolar Chern insulators. 
Instead of carrying a finite chiral edge current of charges, a dipolar Chern insulator has a finite chiral edge current of dipoles with a vanishing net charge current.  

\begin{figure}
    \centering
    \includegraphics[width=0.49\textwidth]{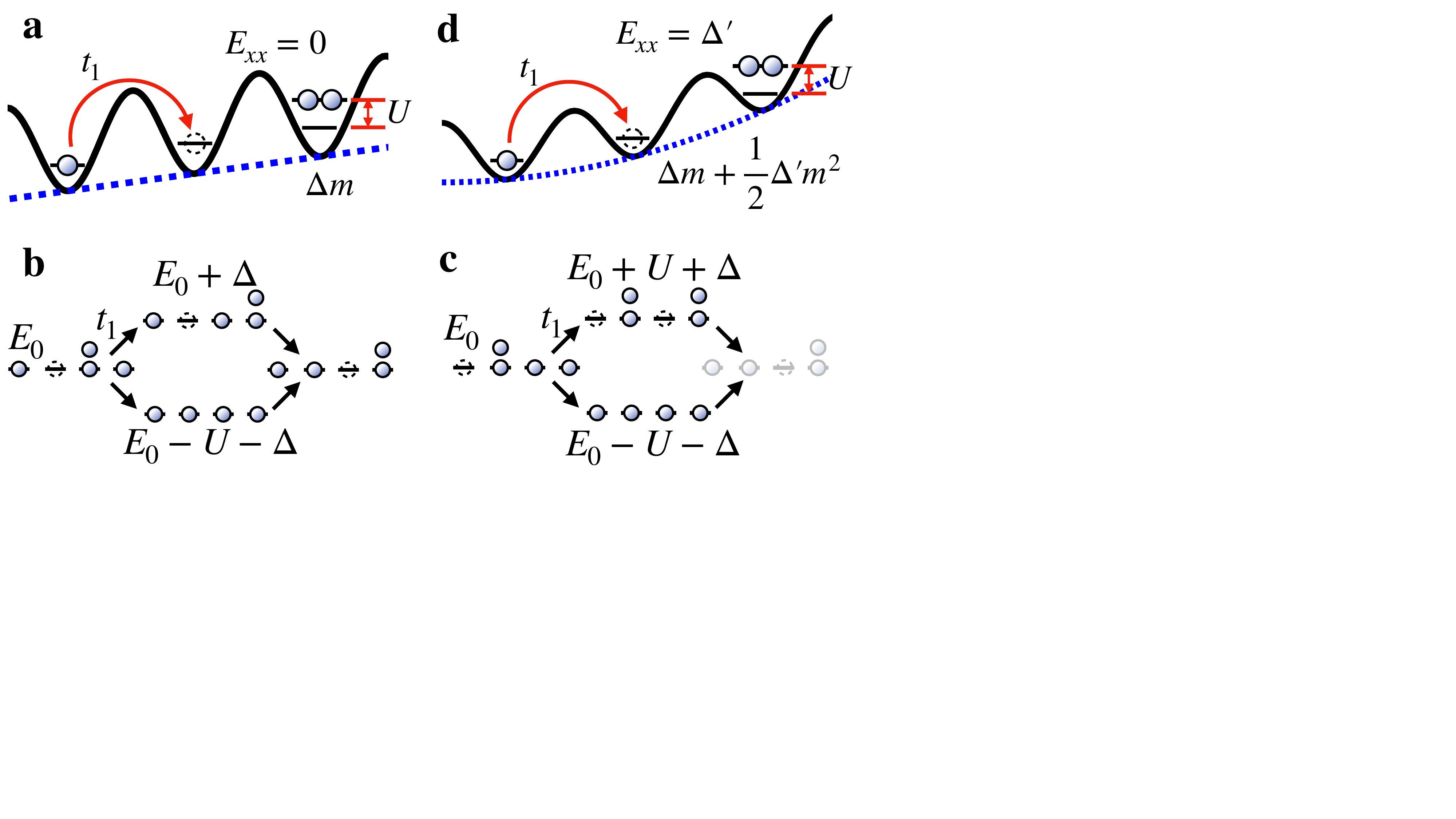}
    \caption{(a) Tensor gauge fields are absent in a linearly tilted lattice. The black curve represents the total external potential. The blue dashed line represents the linear potential. $t_1$ is the single particle tunneling strength and $U$ denotes the onsite interaction.
    (b) Two pathways of second-order processes give rise to the nearest neighbor tunneling of a dipole. $E_0$ denotes the energy of the initial state. The energies of two intermediate states are $E_0+\Delta$ and $E_0-U-\Delta$, respectively.
    (c) The long-range tunneling of a dipole is suppressed due to a destructive interference of two pathways. 
    (d) The blue dashed curve represents the linear potential plus a quadratic potential, which produces a finite rank-2 electric field $E_{xx}$. }
    \label{fig1}
\end{figure}

We start from lineons confined in one dimension. 
As shown in Fig.~\ref{fig1}(a), the Hamiltonian of 1D bosons in a linearly tilted lattice is written as $\hat{H}_B=\hat{K}_B+\hat{V}_B+\sum_m \frac{U}{2}\hat{n}_m(\hat{n}_m-1)$, where $\hat{K}_B=-\sum_{m}(t_{1}\hat{b}^{\dagger}_{{ m}}\hat{b}_{ m+1}+h.c.)$, 
\begin{eqnarray}
  &\hat{V}_B&=\sum_m m\Delta \hat{n}_m, \label{VL}
\end{eqnarray}
and $\hat{n}_m=\hat{b}^{\dagger }_{{ m}}\hat{b}_{{ m}}$.  
$\Delta$ is a constant and $t_1$ is the single-particle tunneling strength.
In the limit where $t_1\ll U$, double occupancy is prohibited, and each site is filled by one boson.  
When $t_1\ll \Delta$ is also satisfied, the tunneling of either the particle or the hole is suppressed. 
Nevertheless, a dipole can hop through some second-order processes. 
As shown in Fig.~\ref{fig1}(b), the second-order process includes two different pathways, which correspond to two different intermediate states with energy $E_0+\Delta$ and $E_0-U-\Delta$ respectively, where $E_0$ is energy of the initial state (Supplementary Materials).
These two pathways yield a finite tunneling amplitude of the dipole,
\begin{equation}
    \hat{\tilde{K}}_L=-\sum_{m}(t_{2}\hat{b}^{\dagger 2}_{{ m}}\hat{b}_{ m+1}\hat{b}_{ m-1}+h.c.),\label{Kd0}
\end{equation}
where $t_2=-t_1^2(\frac{1}{U+\Delta}-\frac1{\Delta})\approx t_1^2U/\Delta^2$. 
Here, we consider the limit $U\ll \Delta$ to avoid the irrelevant tilting-assisted resonance when $U\sim\Delta$. 
If we define $\hat{D}_m^\dagger=\hat b^\dagger_m\hat b_{m-1}$, $\hat{\tilde{K}}_L=-\sum_{m}(t_{2} \hat{D}_m^\dagger \hat{D}_{m+1}+h.c.)$, describing nearest neighbor tunnelings of a dipole. 
Long-range tunnelings vanish due to destructive interference as shown in Fig.~\ref{fig1}(c). 

A more rigorous scheme than the above perturbative approach is using the rotating frame to eliminate $\hat{V}_{B}$. 
The Hamiltonian becomes time-dependent, 
$\hat{\tilde{H}}_{B}=\hat{\mathcal{U}}\hat{H}\hat{\mathcal{U}}^{-1} +i\hbar(\partial_t \hat{\mathcal{U}}) \hat{\mathcal{U}}^{-1}=\hat{\tilde{K}}_{B}(t)+\frac{U}{2}\sum_m\hat{n}_m(\hat{n}_m-1)$,
where $\hat{\tilde{K}}_B=-\sum_m(t_1e^{-i\Delta t/\hbar}\hat b_m^\dag b_{m+1}+h.c.)$, and $\hat{\mathcal{U}}=e^{i\hat{V}_Bt/\hbar}$. We can then derive the Floquet Hamiltonian~\cite{Dalibard2014},
\begin{equation}
    \hat{\tilde{H}}_{L}=\sum_m \bigg(\frac{U}{2}-2t_2\bigg)\hat{n}_m(\hat{n}_m-1)+4t_2\hat{n}_{m+1}\hat{n}_m+ \hat{\tilde{K}}_L.
\end{equation} 

To realize a tensor gauge field for the lineon, we need to add a phase to $t_2$ in Eq.(\ref{Kd0}), i.e., $t_2\rightarrow t_2e^{-i A_{xx}}$. 
Since a constant $A_{xx}$ can be gauged away, here, we consider the simplest means to obtain a time-dependent $A_{xx}$ by adding an extra small quadratic potential to the Hamiltonian, $\hat{H}_B'=\hat{H}_B+\hat{V}_B'$, where
\begin{equation}
\hat{V}_B'=\sum_m \frac{1}{2}m^2\Delta' \hat{n}_m,\label{VQ}
\end{equation}
and $\Delta'$ is a constant. 
To eliminate $\hat{V}_{B}+\hat{V}_B'$, the unitary transformation $\hat{\mathcal{U}}$ needs to be modified to $\hat{\mathcal{U}}'=e^{i(\hat{V}_B+\hat{V}_B')t/\hbar}$.
Using this $\hat{\mathcal{U}}'$, $\hat{\tilde{K}}_B$ is replaced by $\hat{\tilde{K}}'_B=-\sum_m(t_1e^{-i(\Delta+\Delta'(m+1/2))t/\hbar}\hat b_m^\dag \hat b_{m+1}+h.c.)$.
Applying the same method as before, $\hat{\tilde{K}}_L$ is replaced by $\hat{\tilde{K}}'_L=-\sum_m(t_2e^{-i\Delta't/\hbar}\hat b_m^{\dag 2}\hat b_{m+1}\hat b_{m-1}+h.c.)$ in the limit $\Delta\gg \Delta'$.
Compare this expression with Eq.(\ref{Kdp}), we find that $A_{xx}=\Delta' t/\hbar$ is a rank-2 tensor gauge potential varying linearly with time. 

A time-dependent $A_{xx}$ produces a rank-2 electric field, 
\begin{equation}
    E_{xx}=-\frac{\partial A_{xx}}{\partial t}=-\Delta'/\hbar.\label{Exx}
\end{equation}
This result can also be understood from that a quadratic potential in Eq.(\ref{VQ}) produces a linear potential for a dipole. 
From Eq.(\ref{VQ}), we see that when a dipole, i.e., a particle at site $m$ and a hole at site $m-1$, move by one lattice space, the potential energy increases by a constant $\Delta'$. 
A dipole thus experiences a linear potential or equivalently, a constant rank-2 electric field given by Eq.(\ref{Exx}). 

Though a linear potential and a quadratic one is equal to a harmonic trap that merely shifts the origin, the central part of a harmonic trap should be avoided, since the slow increase of the potential there cannot suppress single-particle tunnelings. 
The potential gradient should be strong enough to prohibit single-particle tunnelings. 

\begin{figure}
    \centering
    \includegraphics[width=0.495\textwidth]{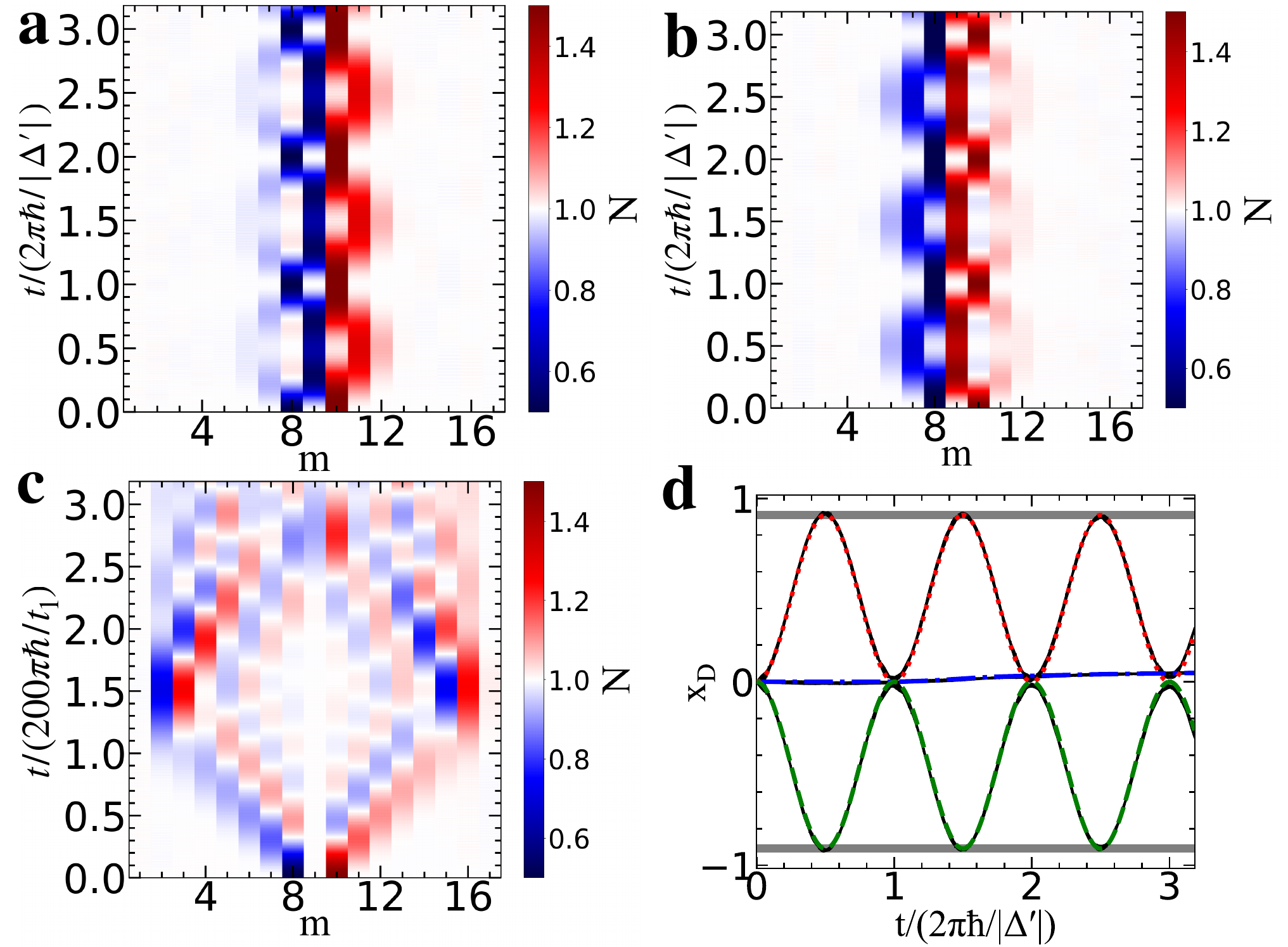}
    \caption{
        Time-dependent density distribution $\langle \hat{n}_m(t)\rangle$ of a lineon.  
    The parameters are chosen as $U/t_1=4$, $\Delta/t_1=40$, and $\Delta'/t_1=-0.01$ (a), $0.01$ (b),$0$ (c). 
    These conditions lead to Bloch oscillations in opposite directions in (a) and (b), and a symmetric expansion in (c). 
   (d) $x_D$ as a function of time. 
   Results of the exact diagonalization of the effective Hamiltonian of a single dipole are shown in red dotted (a), green dashed (b) and blue dash-dotted lines (c). 
   Black curves are results of the full Hamiltonians using TEBD. 
   Horizontal lines denote $-4t_2/\Delta'$.
    }
    \label{fig2}
\end{figure}

To explicitly show the effect of $E_{xx}$, we have numerically computed the dynamics of a lineon using the time-evolution block-decimation (TEBD) method. 
The Hamiltonian is chosen as $\hat{H}_B'$ which includes both a strong linear potential and a weak quadratic potential. 
The initial state includes a single dipole, i.e., a single particle-hole excitation in a Mott insulator. 
Whereas the initial state of the dipole can be an arbitrary wavepacket, for convenience, we let the superposition state of the dipole occupy two lattice sites, i.e., $\ket{\Psi(t=0)}=(\hat{D}^\dagger_{9}+\hat{D}^\dagger_{10})|\text{MI}\rangle/2$, where $|\text{MI}\rangle=\prod_m\hat{b}^\dagger_m|0\rangle$ is a Mott insulator with one particle per site. 
Fig.~\ref{fig2}(a-c) show that the dynamics induced by the rank-2 electric field $E_{xx}=-\Delta'/\hbar$ are distinct from that in the absence of such a field. 
When $\Delta'=0$, a linear external potential for single-particles produces a rank-1 electric field. 
Whereas such a field affects single-particle dynamics, it corresponds to a flat potential for a particle-hole pair.  
As such, the wavepacket of a dipole simply expands symmetrically towards both directions with increasing $t$ as shown in Fig.~\ref{fig2}(c). 
Once $\Delta'\neq 0$, $E_{xx}$  becomes finite.  
As shown by Fig.~\ref{fig2}(a), a  positive $E_{xx}$ produced by a negative  $\Delta'$  and pushes the wavepacket  of a dipole towards right when $t$ increase from zero. 
As $t$ continuously increases, the wavepacket of a dipole oscillates back and forth, i.e.,  a Bloch oscillation of a dipole arises in the presence of a finite $E_{xx}$. 
Changing the sign of $\Delta'$ leads to a negative $E_{xx}$ such that the dipole first moves towards left, as shown in Fig.~\ref{fig2}(b). 

As previously explained, $E_{xx}$ produces a linear external potential for a particle-hole pair, and the effective Hamiltonian of a single dipole is written as $\hat{H}_{\rm eff}=\hat{\tilde{K}}_L-\hbar E_{xx}\sum_m m \hat{D}^\dagger_m \hat{D}_m{/2}$. 
The period of the Bloch oscillation of a dipole is given by $T_D=2\pi/|E_{xx}|$. 
Using Eq.(\ref{Exx}), we obtain  
\begin{equation}
    T_D=2\pi\frac{\hbar}{|\Delta'|}.
\end{equation}
Similarly, the oscillation amplitude is given by $-4t_2/\Delta'$ (Supplementary material). 
For comparison, if the separation between the particle and the hole is larger than one lattice spacing, each of them becomes an isolated excitation. Neither the particle nor the hole moves (Supplementary material). 

To further characterize effects of $E_{xx}$, we consider the change of multipoles in the systems. 
A vector (rank-1) electric field produces a charge current, which changes the dipole moment of the system $P=\sum_{m=1}^L m \langle \hat n_m\rangle$. 
In sharp contrast, a rank-2 electric field does not create a charge current and $P$ remains unchanged. 
Instead, a dipole current is induced by a finite $E_{xx}$ that changes the quadrupole moment $Q=\sum_{m=1}^L m^2 \langle \hat n_m\rangle$. 
We define 
\begin{equation}
  {x}_{D}=\frac1{2}\langle \sum_{m=1}^L[ m-(L{+}1)/2]^2\hat n_m\rangle-\frac{1}{24} L (L^2-1).
\end{equation}
In the absence of a dipole, $\langle\hat{n}_m\rangle=1$, $x_D=0$. 
When there is a doubloon at site $m+1$ and a hole at site $ m$, $ x_D=m-L/2$. 
We therefore read from ${x}_D$ the position of the dipole.  

It is clear that ${x}_{D}=\frac{1}{2}[Q-(L+1)P]+\frac{1}{8}(L+1)^2N-\frac{1}{24}L(L^2-1)$, where $N=\sum_{m=1}^L \langle \hat n_m\rangle$ is the total particle number. 
Since both $P$ and $N$ are conserved (Supplementary Materials), ${x}_D$ directly reflect the change of the quadrupole moment due to $E_{xx}$. 
Experimentalists thus need to measure $N$, $P$ and $Q$ to unfold the effect of $E_{xx}$.  
Fig.~\ref{fig2}(d) shows that $x_D$ oscillates as time goes by when $\Delta'\neq 0$.  
The results of $H_{\rm eff}$ agree well with the numerical solution of the full Hamiltonian of bosons $\hat{H}'_B$ using TEBD. 
We have smoothed the curves by averaging the results within a much smaller time scale $2\pi\hbar/\Delta$ (Supplementary Materials).  Unlike the conventional Bloch oscillation that changes $P$, the dipolar Bloch oscillation periodically modulates $Q$ while $P$ remains unchanged.

\begin{figure}
    \centering
    \includegraphics[width=0.48\textwidth]{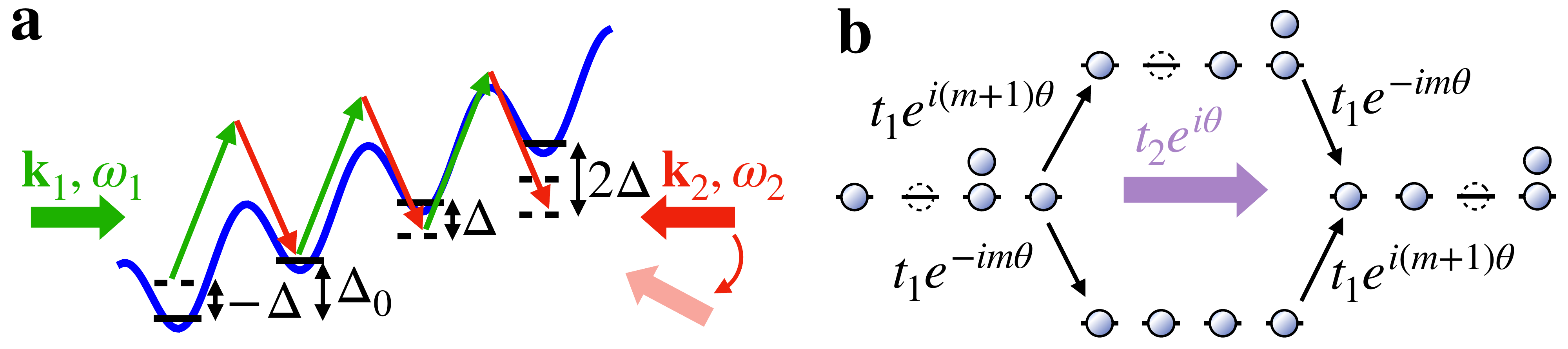}
    \caption{(a) Raman lasers (green and red arrows) produce photon-assisted tunnelings in a linearly tilted lattice. 
    A finite detuning $\Delta=\Delta_0-\hbar\omega$ leads to a residual linear potential. 
    (b) The site-dependent phase of the single-particle tunneling provides a phase in the tunneling of a dipole (violet arrow).
    }
    \label{fig3}
\end{figure}

An alternative means is to realize $A_{xx}$ directly. 
As shown in Fig.~\ref{fig3}, photon-assisted tunnelings that carry the phases of external lasers can be made position-dependent~\cite{Bloch2013, Ketterle2013}. 
Instead of a resonant coupling, we consider a finite detuning, $\Delta=\Delta_0-\hbar\omega$, where $\Delta_0$ is the energy mismatch between adjacent sites of the linearly tilted lattice, and $\omega=\omega_1-\omega_2$ is the difference between frequencies of the two Raman lasers.  
As such, there exists a residual linear potential described by Eq.(\ref{VL}).
The one-dimensional system is then described by the same Hamiltonian described before. 
Most of the remaining discussions apply except an important difference that $\hat{K}_B$ is replaced by $\hat{K}'_B=-\sum_{m}(t_{1}e^{-im\theta}\hat{b}^{\dagger}_{{ m}}\hat{b}_{ m+1}+h.c.)$.
Correspondingly, the difference in the phase carried by the particle and the hole leads to a finite phase in the tunneling of a dipole. 
The kinetic energy of a dipole is then readily given by Eq.(\ref{Kdp}), where $A_{xx}=\theta$, and there is no need to introduce an extra quadratic potential. 
To have a time-dependent $A_{xx}$, we recall that $\theta=({\bf k}_1-{\bf k}_2)\cdot \bf{a}$, where ${\bf k}_{i=1,2}$ is the wave vectors of the Raman lasers and ${\bf a}$ is the unit vector of the lattice. 
The incident angles of the Raman lasers can then be changed as time goes by. 

Whereas the previously discussed method works for lineons, the tensor gauge fields for planons need separate discussions. 
Unlike lineons that are confined in 1D, planons could move in two directions that are perpendicular to the dipole moment. 
As such, more sophisticated designs are required to deliver a higher dimensional system subject to tensor gauge fields.
A ring-exchange interaction allows a planon to move, 
\begin{equation}
\hat{\tilde{K}}_{P}=-\sum_{\bf m}(t_2'\hat{b}^\dagger_{\bf m}\hat{b}^\dagger_{{\bf m}+\hat{\bf x}+\hat{\bf y}}\hat{b}_{{\bf m}+\hat{\bf x}}\hat{b}_{{\bf m}+\hat{\bf y}}+h.c.),
\end{equation}
where ${\bf m}=(m_x,m_y)$ is the lattice index and $\hat{\bf x}$ and $\hat{\bf y}$ are the unit vectors in the $x$ and $y$ directions, respectively. 
If we define the creation operator for a planon as $\hat{D}^\dagger_{{\bf m},{\bf \hat{n}}}=\hat{b}^\dagger_{\bf m}\hat{b}_{{\bf m}-\hat{\bf n}}$, where $\hat{\bf n}=\hat{\bf x}, \hat{\bf y}$, $t_2'$ corresponds to the tunneling strength of a planon. 
When applying similar second-order processes of single-particle tunnelings as that in discussions about lineons, it produces both the ring-exchange interaction and kinetic energies of lineons (Supplementary Materials).  

\begin{figure}
    \centering
    \includegraphics[width=0.495\textwidth]{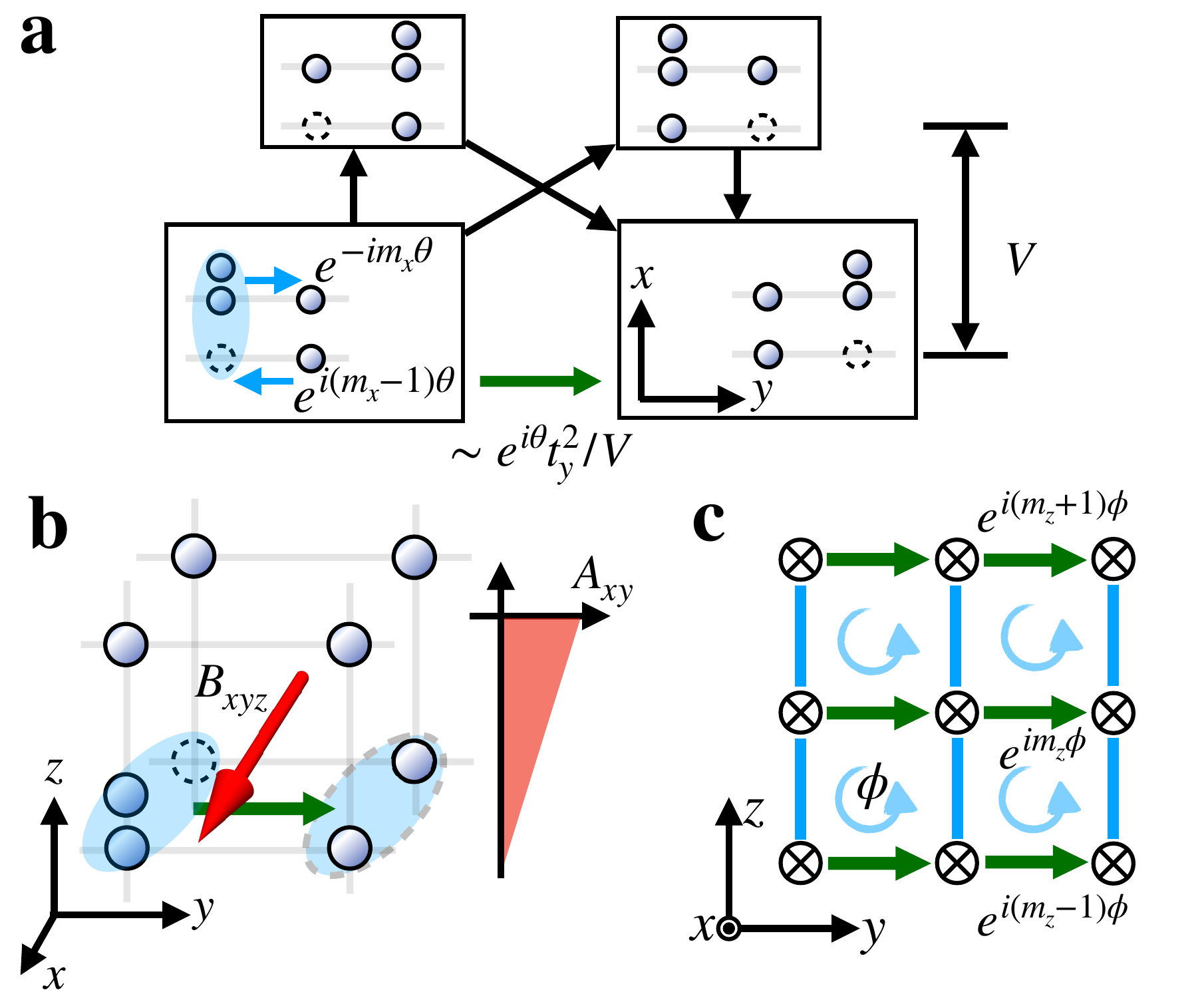}
    \caption{
        (a) The tunneling of a planon (green thick arrow) through second order processes of single-particle tunnelings (thin black arrow). 
        (b) A $z$-dependent $A_{xy}$ provides a magnetic field for the planon. 
        (c) The side view of (b) represents a dipolar Harper-Hofstadter model where each plaquette is threaded with a finite flux $\phi$. 
    }
    \label{fig4}
\end{figure}

To access purely a ring-exchange interaction, and more importantly, to imprint a controllable phase to the ring-exchange interaction such that the tensor gauge field for a planon becomes finite, we consider an alternative scheme using particles with long-range interactions in a lattice. 
As shown in Fig.~\ref{fig4}(a), a strong lattice potential in the $x$ direction suppresses tunnelings along this direction, i.e., $t_x=0$. 
In the $y$ direction, Raman lasers imprint a $x$-dependent phase in the tunneling, $t_ye^{i m_x\theta}$. 
The Hamiltonian is written as $\hat{H}_M=\hat{K}_M+\hat{U}_M$, where
\begin{eqnarray}
&\hat{K}_M&=-\sum_{\bf m} (t_ye^{im_x\theta}\hat{b}^\dagger_{\bf m}\hat{b}_{{\bf m}+\hat{\bf y}}+h.c.),\\
&\hat{U}_{M}&=\frac{U}{2}\sum_{\bf m}\hat{n}_{\bf m}(\hat{n}_{\bf m}-1)+V\sum_{\langle{\bf m},{\bf n}\rangle}\hat{n}_{\bf m}\hat{n}_{\bf n}.
\end{eqnarray}
$V$ denotes the nearest neighbor interaction. 
Here, we consider an isotropic nearest-neighbor interaction for simplicity. 
Anisotropic interactions do not change the main results. 

When a dipole is aligned along the $x$ direction, the tunneling of either the particle or the hole is suppressed in the limit where $V\gg t_y$, as such tunnelings lead to an extra energy penalty, as shown in Fig.~\ref{fig4}(a). 
Second-order processes allow such a dipole to tunnel in the $y$ direction. 
The kinetic energy of this planon is written as   
\begin{equation}
\hat{K}_{P}=\sum_{\bf m}(J_ye^{-iA_{xy}}\hat{D}^\dagger_{\bf m,\hat{\bf x}}  \hat{D}_{{\bf m}+\hat{\bf y},\hat{\bf x}}+h.c.),
\end{equation}
where $J_y=-2t_y^2/V$, and $A_{xy}=\theta$. 
A time-dependent $ A_{xy}$ gives rise to an electric field $E_{xy}=-\partial\theta/\partial t$ acting on a planon, which
performs Bloch oscillations in the direction perpendicular to the dipole moment. 

In 3D, the variance of $A_{xy}$ in the $z$~direction leads to a magnetic field $B_{xyz}=- \partial A_{xy}/\partial z$. 
As shown in Fig.~\ref{fig4}(b), such a magnetic field can be accessed by introducing a $z$-dependent $\theta=\phi z$. 
To this end, we consider a three-dimensional model, $\hat{H}'_M=\hat{K}'_M+\hat{U}_M$,
where 
\begin{equation}
 \hat{K}'_M =-\sum_{\bf m} (t_ye^{im_xm_z\phi}\hat{b}^\dagger_{\bf m}\hat{b}_{{\bf m}+\hat{\bf y}}+t_z\hat{b}^\dagger_{\bf m}\hat{b}_{{\bf m}+\hat{\bf z}}+h.c.).\label{Kmp}
\end{equation}
$\hat{m}=(m_x, m_y,m_z)$ has been promoted to the index of a three-dimensional lattice. 
For a fixed $m_x$, the phase carried by $t_y$ changes linearly as a function of $m_z$. 
We thus just need to generalize the scheme of realizing the Harper-Hofstadter model to a $m_x$ dependent one. 
The simplest scheme is to use the synthetic dimension formed by internal degrees of freedom like hyperfine spins or different atomic species, which may feel different vector gauge potentials. For instance, a seminal experiment realized opposite vector gauge potentials for two spins \cite{Bloch2013}. To realize the dipolar Harper-Hofstadter model, experimentalists just need to turn on a strong interaction between these two spin components.
As the tunneling in the $x$ direction is suppressed, it is sufficient to consider only two layers with $m_x=0$ and $m_x=1$, respectively. 
In other words, only a two-component system is required. 
Applying component-selective Raman lasers that do not affect the first component, the tunneling of this component is unchanged. 
$t_y$ in the first layer with $m_x=0$ thus does not acquire an additional phase. 
In contrast, the Raman lasers imprint a phase to the tunneling of the second component. 
$t_y$ in the second layer with $m_x=1$ thus becomes $t_ye^{im_z\phi}$. 
The minimal version of Eq.(\ref{Kmp}) is then realized. 
Alternatively, if one directly implements the scheme in~\cite{Bloch2013}, where these two components feel opposite vector gauge potentials, $\phi\rightarrow 2\phi$.

Once accessing Eq.(\ref{Kmp}) in laboratories, a dipolar Harper-Hofstadter model is realizable. 
Similar to the previous discussions, a dipole aligned in the $x$ direction can also tunnel in the $z$ direction with a tunneling strength $J_z=-t_z^2/V$. 
If we view the system along the $x$-axis, the dipole appears to be a two-dimensional particle moving in the $y-z$ plane, as shown in Fig.~\ref{fig4}(c).
This thus realizes a dipolar Harper-Hofstadter model,
\begin{equation}
   \begin{split}
        \hat H_{\rm dHH}=&\frac{1}{2}\sum_{\bf m}(J_y e^{-im_z \phi}\hat D_{\bf m,\hat x} \hat D^\dag_{\bf m+\hat y,\hat x}\\
        &+J_{z}\hat D_{\bf m,\hat x} \hat D^\dag_{\bf m+\hat z,\hat x}+h.c.).
   \end{split}
\end{equation}
where the flux included in a single plaquette becomes finite. 
This dipolar Harper-Hofstadter model hosts many new quantum phenomena. For instance, 
when the bulk is gapped, the state can be referred to as a dipolar Chern insulator. Unlike conventional Chern insulators that feature chiral charge current, here, the net charge current at the edge of a dipolar Chern insulator is zero and the topological transport of the system is provided by the chiral dipole current at the edges. 

We have shown that some simple protocols fulfill the lofty goal of creating synthetic tensor gauge fields in laboratories. We have used the rank-2 scalar charge theory \cite{Pretko2017_2}, which satisfies the gauge transformation, $A_{ij}\rightarrow A_{ij} +\partial_i\partial_j\alpha$ where $\alpha$ is a scalar field, as an example.  
Our scheme can be generalized to implement other tensor gauge theories in laboratories. 
It is also possible to realize even higher rank gauge fields using related schemes.  
We hope that our work may stimulate more interest to study higher rank gauge fields.  

This work is supported by the U.S. Department of Energy, Office of Science through the Quantum Science Center (QSC), a National Quantum Information Science Research Center and National Science Foundation (NSF) through Grant No. PHY-2110614. S.Zhang is supported by National Natural Science Foundation of China(Grant No.12174138).

\onecolumngrid

\newpage 
\vspace{0.2in}

\centerline{\bf Supplementary Material}

\vspace{0.1in}

In this supplementary material, we present results of the fast oscillation due to the micromotion of single particles, tunneling amplitude of dipoles from perturbation method, the amplitude of Bloch oscillation, immobility of an isolated particle or hole, 
changes in the dipole and quadrupole moments due to a rank-2 electric field, and a scheme that produces both the ring-exchange interaction and the kinetics of lineons.

\subsection{Fast oscillations due to the micromotion of single-particles}
\begin{figure}[h]
    \centering
    \includegraphics[width=0.69\textwidth]{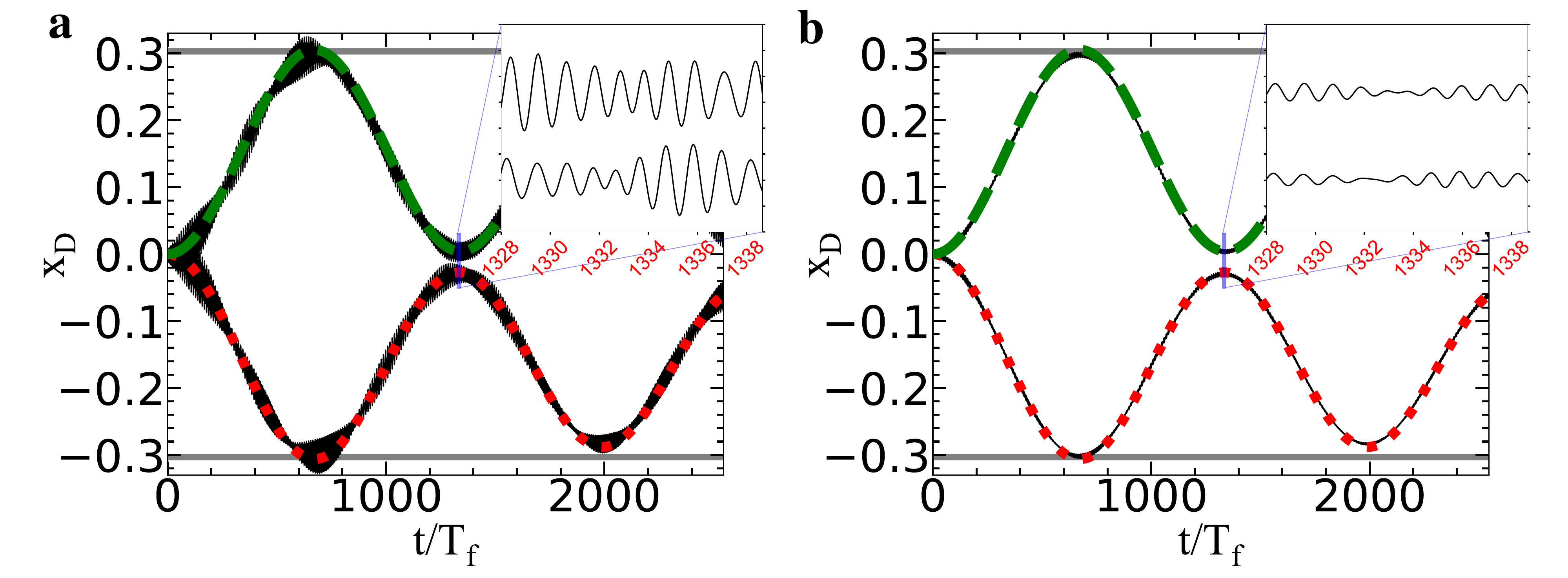}
    \caption{
    $x_D$ as a function of the time before (a) and after (b) averaging over a small time period $T_f=2\pi\hbar/\Delta$.  The parameters are chosen as $U/t_1=4$, $\Delta/t_1=40$ and
   $\Delta'/t_1=-0.03$, $0.03$. Results of the exact diagonalization of the effective Hamiltonian of a single dipole are shown in red dotted ($\Delta'/t_1=0.03$), green dashed ($\Delta'/t_1=-0.03$) lines. Black curves are the results of the exact diagonalization of the full Hamiltonians. Horizontal lines denote $-4t_2/\Delta'$. Insets show the results in a small time window. 
   }
    \label{fig:s0}
\end{figure}

When deriving the effective Hamiltonian for the dipole using the Floquet theory, we have dropped off the micromotion of single-particles. In the presence of a strong linear potential, a single particle experiences a Bloch oscillation with a frequency $\omega_s=\Delta/\hbar$. 
This is a micromotion in a much smaller time scale than that of the Bloch oscillation of the dipole. 
Whereas numerical results clearly show such fast oscillations, we  consider the average over time in such a small scale $T_f=2\pi/\omega_s$ to smooth the results. 
Here, we show similar numerical results as that in Fig.~2 (e) using exact diagonalization with $L=7$. The main figure in Fig.~\ref{fig:s0}(a) show the result before doing the average. 
The fast oscillations are highlighted in the inset. 
Fig.~\ref{fig:s0}(b) shows the result after doing the average.

\subsection{Tunneling amplitude of dipoles from perturbation method}

In main text, we have obtained the effective Hamiltonian in Eq.(4) using the Floquet theory. Here, we provide a detailed derivation of the effective Hamiltonian using perturbation method. 
We consider the original Hamiltonian $\hat{H}_B=\hat{H}_0+\hat{H}_1$ where
\begin{equation}
\hat{H}_0=\sum_mm\Delta\hat{n}_m+\frac{U}{2}\sum_m\hat{n}_m(\hat{n}_m-1).
\end{equation}
The single-particle tunneling
\begin{equation}
\hat{H}_1=-t_1\sum_m(\hat{b}^\dag_m\hat{b}_{m+1}+h.c.)
\end{equation}
can be considered as a perturbation in the limit $\Delta, U\gg t_1$. As discussed in main text, we define the states as $|m\rangle=|1\cdots 1021\cdots 1\rangle$ where the particle number at the $m$th and $(m+1)$th sites are $0$ and $2$, respectively. The particle number is $1$ at any other site. These orthogonal states are the eigenstates of $\hat{H}_0$ with the same eigenenergy $E_0$, which are used as the basis states in the perturbation approach. 
The matrix elements $M_{n,m}=M_{m,n}^*$ of the perturbation method are written as 
\begin{equation}
M_{n,m}=\sum_\alpha\frac{\langle n|\hat{H}_1|\alpha\rangle\langle\alpha|\hat{H}_1|m\rangle}{E_0-E_\alpha},
\label{tme}
\end{equation}
where $|\alpha\rangle$ are the intermediate states with eigenenergy $E_\alpha$. When $m=n$, the second-order processes provide a finite energy shift which is independent of $m$. If $m<n$,  we need to consider two different cases.

1) $n=m+1$. As shown in Fig.1(b), a particle-hole pair is formed by a hole and a doubloon in two adjacent sites $m$ and $m+1$. In the first pathway, one particle in the doubloon tunnnels to the right, creating an intermediate state $|\alpha_1\rangle=|1\cdots 10121\cdots 1\rangle$ with energy $E_1=E_0+\Delta$, where the $m$th and $(m+2)$th sites have $0$ and $2$ particles respectively. Then the hole moves to the right, resulting in a final state $|m+1\rangle$ with the same energy as the initial state $|m\rangle$. The particle-hole pair thus hops by one lattice spacing. In the second pathway,  the hole moves to the right, annihilating one of the two particles in the doubloon, and the intermediate state is $|\alpha_2\rangle=|11\cdots 1\rangle$ with energy $E_2=E_0-U-\Delta$. The remaining particle in the doubloon then tunnels to the right, creating the same final state as the first pathway. These two pathways yield a finite transition matrix element in Eq.(\ref{tme}) as
\begin{equation}
M_{m+1,m}=\frac{(-t_1)(-2t_1)}{-\Delta}+\frac{(-\sqrt{2}t_1)(-\sqrt{2}t_1)}{U+\Delta}=-2t^2_1\frac{U}{\Delta(\Delta+U)}.
\end{equation}
Using the effective Hamiltonian in Eq.(3) in main text, such a matrix element gives rise to a finite tunneling amplitude of the dipole, 
\begin{equation}
-2t_2=\langle m+1|\hat{\tilde{K}}_L|m\rangle =-t_2\langle m+1|\big\{\sum_n\hat{b}^{\dag 2}\hat{b}_{m+1}\hat{b}_{m-1}+h.c.\big\}|m\rangle. 
\end{equation}
Comparing the above two equations, we obtain
\begin{equation}
t_2=t^2_1\frac{U}{\Delta(\Delta+U)}\approx\frac{t^2_1 U}{\Delta^2}.
\end{equation}
The last approximation is implemented when $U\ll\Delta$.

2) $n-m>1$. As shown in Fig.1(c), there are also two different pathways. In the second pathway, the hole moves to the right to annihilate one of the two particles in the doubloon and the intermediate state is the same with the above case. In the first pathway, the doubloon remains unchanged but a single particle moves right to create another particle-hole pair, this intermediate state has the energy  $E_1=E_0+U+\Delta$ instead of $E_0+\Delta$. The matrix element in Eq.(\ref{tme}) can be calculated using the similar method. We obtain
\begin{equation}
M_{n,m}=\frac{(-t_1)(-2t_1)}{-U-\Delta}+\frac{(-\sqrt{2}t_1)(-\sqrt{2}t_1)}{U+\Delta}=0,
\end{equation}
which means that the contribution of two pathways are cancelled with each other. The long-range tunnelings of a dipole is thus prohibited.

\subsection{The amplitude of Bloch oscillation}
Within the effective theory, the dipole dynamics is captured by a tight-binding Hamiltonian with a linear potential,
\begin{equation}
    \hat H_{\rm eff}=\sum_{m}-2t_2(\ket{m}\bra{m+1}+h.c.)+\Delta'm\ket{m}\bra{m},
\end{equation}
where $\ket{m}=\hat D^\dag_m\ket{\rm MI}/\sqrt{2}$ denotes the normalized single dipole state.
The eigenstates of the effective Hamiltonian $\hat H_{\rm eff}$ are  the Wannier Stark states. The time evolution operator is written as \cite{Hartmann2004}
\begin{equation}
    U(t)=\sum_{m,m'}J_{m-m'}(8t_2 \sin(\Delta' t/2\hbar)/\Delta')e^{i(m-m')(\pi-\Delta' t/\hbar)/2-im'\Delta' t/\hbar}\ket{m}\bra{m'},
\end{equation}
where $J_m(x)$ is the Bessel function of the first kind.
$U(t)$ has a period $T=2\pi\hbar/\Delta'$. Without loss of generality, we consider an initial state $\ket{\psi(0)}=(\ket{0}+\ket{1})/\sqrt{2}$. The quantum state reaches its maximal extension at half of the oscillation period. The averaged dipole displacement corresponding to the oscillation amplitude is written as
\begin{equation}
    \bra{\psi(T/2)}m\ket{\psi(T/2)}-\bra{\psi(0)}m\ket{\psi(0)}=\sum_{m}\frac1{2}\bigg(J_m(8t_2/\Delta')-J_{m-1}(8t_2/\Delta')\bigg)^2 m-\frac{1}{2}=-\frac{4t_2}{\Delta'}.
\end{equation}

\subsection{Immobility of an isolated particle or hole}
\begin{figure}[h]
  \centering
  \includegraphics[width=0.4\textwidth]{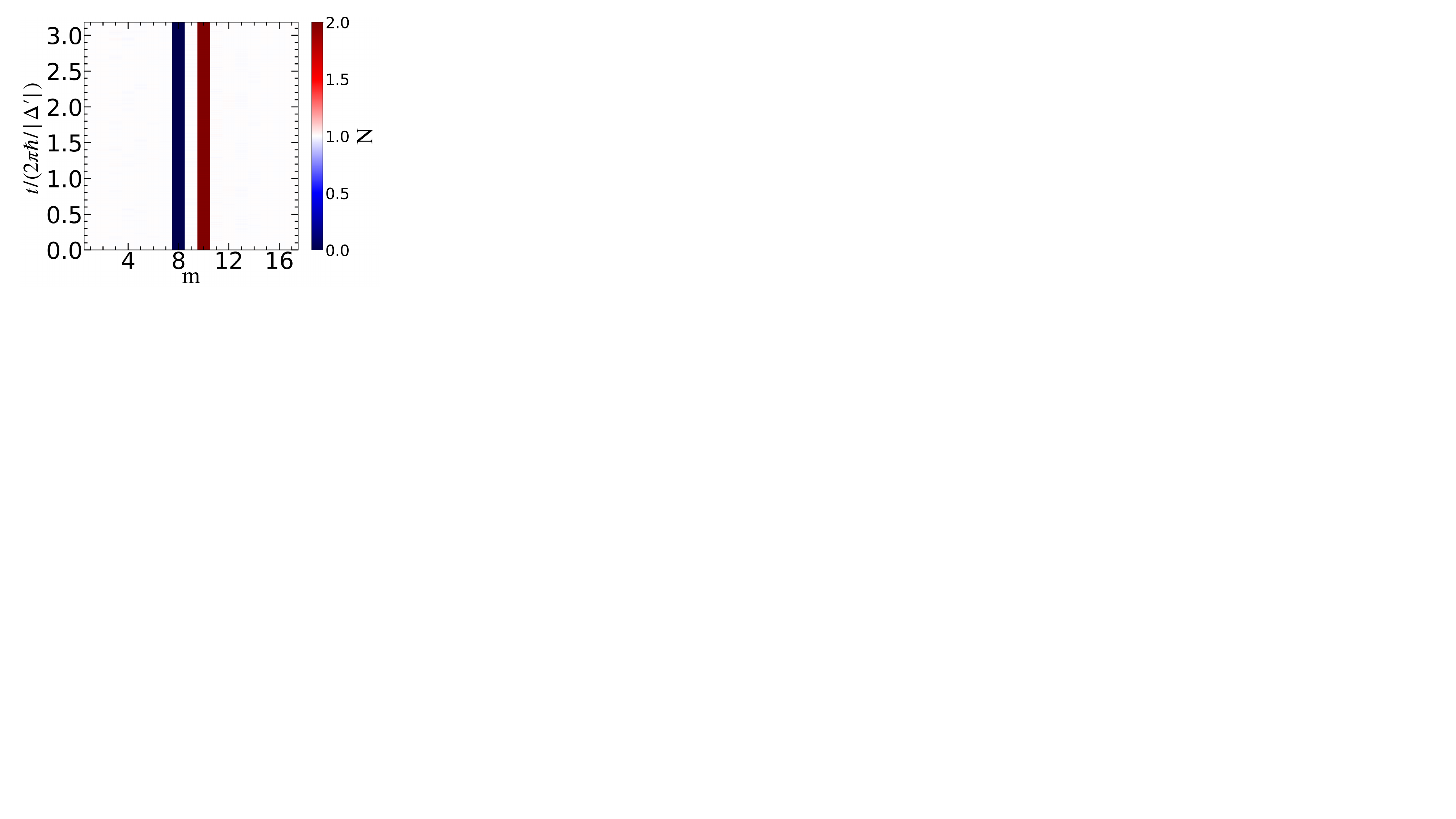}
  \caption{
    Time-dependent density distribution $\langle \hat{n}_m(t)\rangle$.  
    The parameters are chosen as $U/t_1=4$, $\Delta/t_1=40$, and $\Delta'/t_1= 0.01$. Neither the isolated particle or hole moves. 
 }
  \label{fig:s3}
\end{figure}

If the separation between a particle and a hole is larger than one lattice spacing, each of them is an isolated excitation and thus cannot move. 
For instance, if the initial state is $\ket{\Psi(t=0)}=\hat{b}_{8}\hat{b}^{\dagger }_{10}|\text{MI}\rangle/\sqrt{2}$, both the particle and the hole remain stationary when applying the same Hamiltonian in Fig.~2 in the main text, as shown in Fig.~\ref{fig:s3}.

\subsection{Changes in the dipole and quadrupole moments due to a rank-2 electric field}

\begin{figure}[h]
  \centering
  \includegraphics[width=0.4\textwidth]{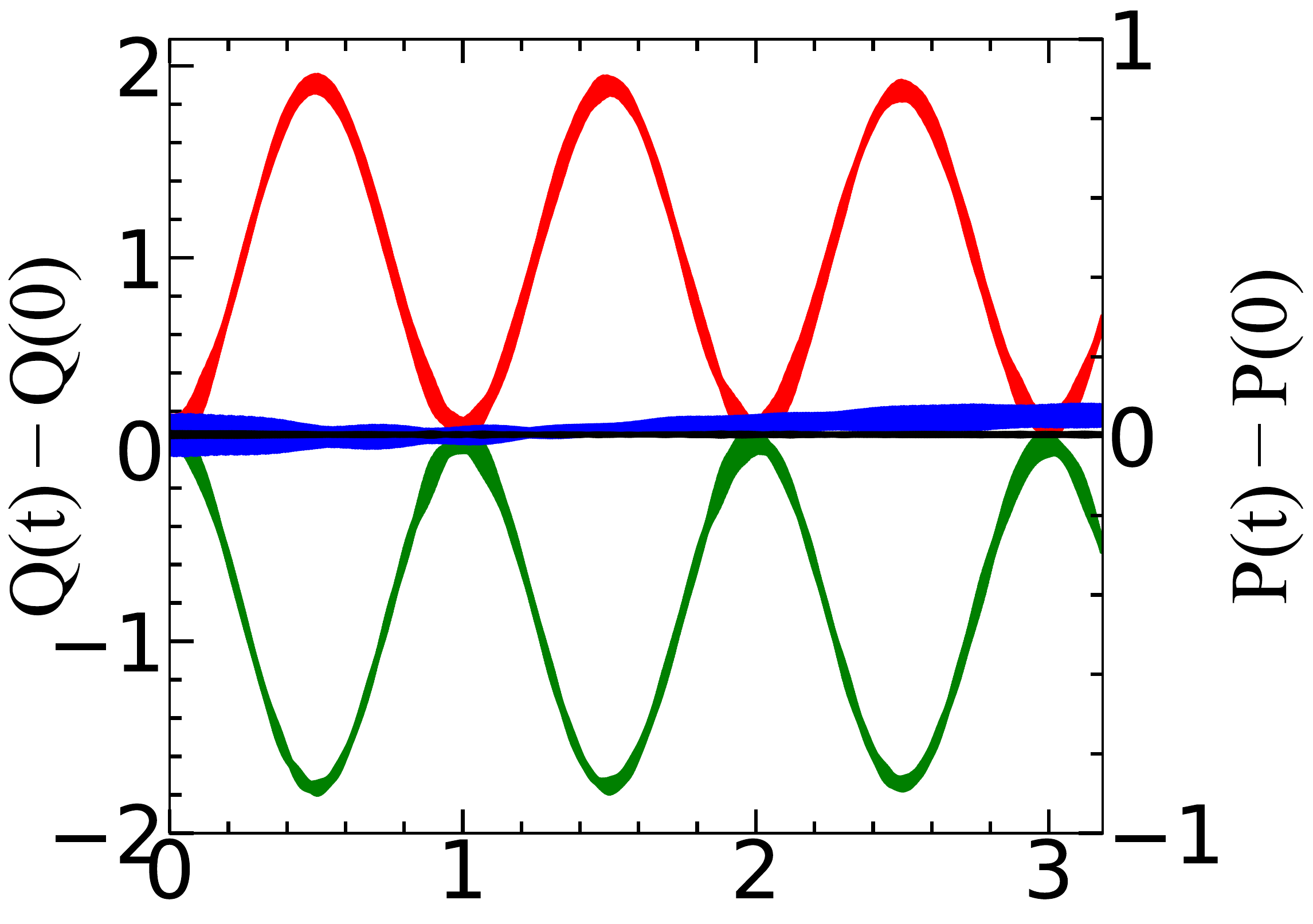}
  \caption{
    Dipole (Quadrupole) moment relative to its initial value $P(t)-P(0)$ ($Q(t)-Q(0)$) as functions of time.
    The parameters are chosen as $U/t_1=4$, $\Delta/t_1=40$. Red, green and blue curves correspond to the evolution of quadrupole moment  when $\Delta'/t_1= -0.01, +0.01, 0$ respectively. Black line shows the evolution of dipole moment in all these three cases.
 }
  \label{fig:s4}
\end{figure}

Distinct from a rank-1 electric field that changes the dipole moment of the system, a rank-2 electric field modifies the quadrupole moment and preserves the dipole moment. In the presence of a constant rank-2 electric field, a dipole undergoes a Bloch oscillation and produces a dipole current. 
As such, the quadrupole moment oscillates. 
Since the particle and hole in a dipole always move in the same direction as a bound state, 
the net charge current is zero and the dipole moment remains constant.
In Fig.~\ref{fig:s4}, we show the time-dependent dipole moment $P(t)=\sum_{m=1}^{L}m\langle \hat n_m(t)\rangle$ and quadrupole moment $Q(t)=\sum_{m=1}^{L}m^2\langle \hat n_m(t)\rangle$ for the same Hamiltonian and initial state as that in Fig.~2 in the main text.
For a finite rank-2 electric field $\Delta'\neq 0$ (red and green curves), the quadrupole moment oscillates periodically. 
When $\Delta'=0$, a vanishing $E_{xx}$ does not change the quadrupole moment, which thus remains constant. 
In practice, there are residual single-particle Bloch oscillations of the particle and the hole in a timescale $\sim 1/\Delta$ much smaller than the time scale of the Bloch oscillation of the dipole $\sim1/\Delta'$. This leads to a deviation of $Q(t)-Q(0)$ from zero (blue curve), which is suppressed by averaging over time. 
With increasing the linear tilting, the amplitude of the single-particle Bloch oscillations of the particle and the hole is suppressed down to zero and such a small deviation of $Q(t)$ from $Q(0)$ vanishes. 
In long times, the wavefunction of the dipole touches the boundary and such a finite size also leads to a small deviation of $Q(t)$ from $Q(0)$. With increasing the size of the system, this small deviation of $Q(t)-Q(0)$ from zero caused by the finite size also vanishes. 
Distinct from the quadrupole moment, the dipole moment depicted by the black curve remains constant for any $\Delta'$.

\subsection{Producing both the ring-exchange interaction and kinetics of lineons}

\begin{figure}
    \centering
    \includegraphics[width=0.69\textwidth]{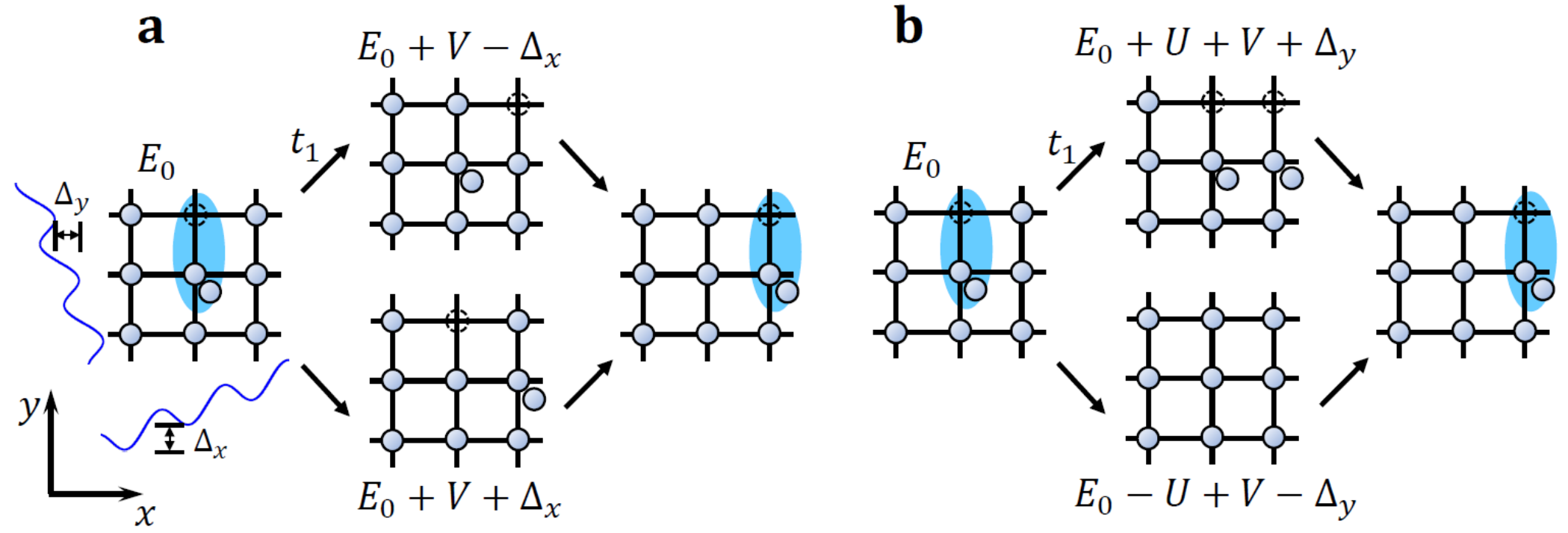}
    \caption{The ring-exchange interaction.(a) and (b) correspond to the second-order processes of single-particle tunnelings along the $x$ and the $y$ directions, respectively. }
    \label{figs1}
\end{figure}

We consider a two-dimensional Hamiltonian that is written as $\hat{H}_B=\hat{K}_B+\hat{V}_B+\hat{U}_B$ where
\begin{eqnarray}
\begin{aligned}
   \hat{K}_B=&-\sum_{{\bf m},\hat{\bf n}}(t_1\hat{b}_{\bf m}\hat{b}_{{\bf m}+\hat{\bf n}}+h.c.),\label{kb0_ring} \\
     \hat{V}_B=&\sum_{\bf m}(m_x\Delta_x+m_y\Delta_y)\hat{n}_{\bf m},\label{vb0_ring} \\
     \hat{U}_B=&\frac{U}{2}\sum_{\bf m}\hat{n}_{\bf m}(\hat{n}_{\bf m}-1)+V\sum_{\bf m}\hat{n}_{\bf m}(\hat{n}_{\bf m+\hat{x}}+\hat{n}_{\bf m+\hat{y}}).\label{ub0_ring}
\end{aligned}
\end{eqnarray}
$\Delta_x$ and $\Delta_y$ are energy matches between nearest-neighbor sites along the $x$ and $y$ directions, respectively. 
In 2D, the finite nearest-neighbor interaction $V$ is necessary for realizing correlated pair tunnelings through second order processes of single-particle tunnelings. 
Such $V$ can be realized by using polar molecules or Rydberg atoms arrays. 
As shown by Fig.\ref{figs1}, when $V\ne 0$, four pathways yield a finite amplitude of the ring exchange interaction, allowing the a planon to tunnel along the perpendicular direction of the dipole moment,
\begin{equation}\label{ring_exchange_term}
  \hat{\tilde{K}}_{P}=-\sum_{\bf m}(t_2'\hat{b}^\dagger_{\bf m}\hat{b}^\dagger_{{\bf m}+\hat{\bf x}+\hat{\bf y}}\hat{b}_{{\bf m}+\hat{\bf x}}\hat{b}_{{\bf m}+\hat{\bf y}}+h.c.),
\end{equation} 
where $t'_2=t'_{2x}+t'_{2y}$ and $t'_{2x}=\frac{2V}{V^2-\Delta^2_x}t^2_1$, $t'_{2y}=\frac{2V}{V^2-(U+\Delta_y)^2}t^2_1$. 

\begin{figure}
    \centering
    \includegraphics[width=0.9\textwidth]{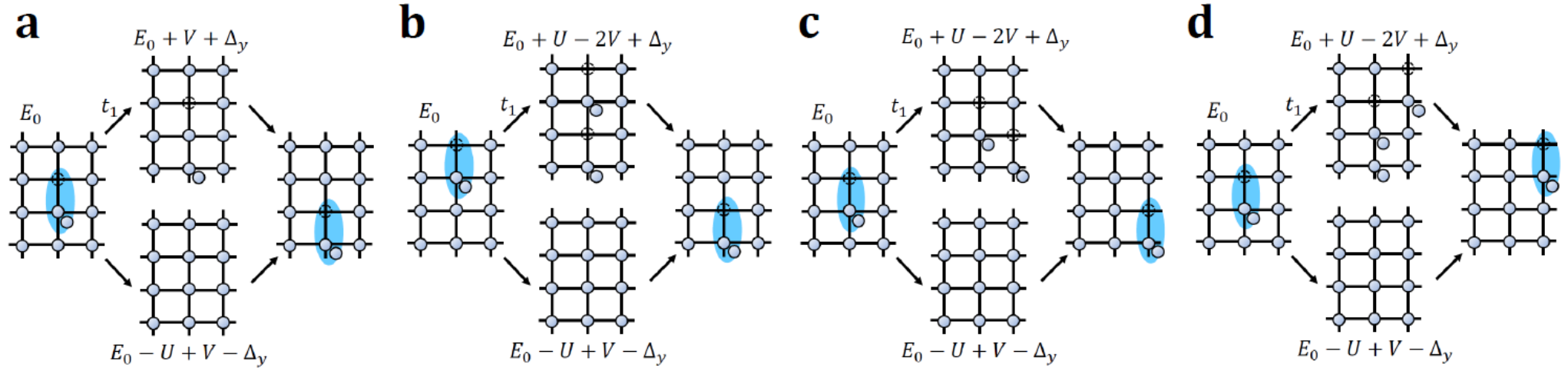}
    \caption{(a) and (b) depict the nearest-neighbor and next-nearest-neighbor hopping of a lineon, respectively.
    (c) and (d) depict the tunnelings of a dipole along the diagonal directions. 
    }
    \label{figs2}
\end{figure}

Some other tunnelings of a dipole can also be produced by second order processes, as shown in Fig.\ref{figs2}. 
For example, the tunneling of a lineon exists,
\begin{equation}\label{corr_tunnel_term}
  \hat{\tilde{K}}_{L}=-\sum_{\bf m}(t_2\hat{b}^{\dag 2}_{\bf m}\hat{b}_{\bf m+\hat{y}}\hat{b}_{\bf m-\hat{y}}+\tilde{t}_2\hat{b}^\dag_{\bf m-\hat{y}}\hat{b}^\dag_{\bf m}\hat{b}_{\bf m+\hat{y}}\hat{b}_{\bf m+2\hat{y}}+h.c.),
\end{equation} 
where $t_2=\frac{(2V-U)}{(V+\Delta_y)(V-U-\Delta_y)}t^2_1$ and $\tilde{t}_2=\frac{V}{(U-2V+\Delta_y)(U-V+\Delta_y)}t^2_1$ correspond to the amplitude of the nearest-neighbor and the next-nearest-neighbor tunneling of a lineon, respectively. Furthermore, the dipole can tunnel along the diagonal direction,
\begin{equation}\label{diag_tunnel_term}
  \hat{\tilde{K}}_\mathrm{diag}=-\sum_{\bf m}(\tilde{t}_{\hat{x}+\hat{y}}\hat{b}^\dagger_{\bf m}\hat{b}_{\bf m+\hat{y}}\hat{b}_{\bf m+\hat{x}+\hat{y}}\hat{b}^\dag_{\bf m+\hat{x}+2\hat{y}}+\tilde{t}_{\hat{x}-\hat{y}}\hat{b}^\dagger_{\bf m}\hat{b}_{\bf m+\hat{y}}\hat{b}_{\bf m+\hat{x}-\hat{y}}\hat{b}^\dag_{\bf m+\hat{x}}+h.c.),
\end{equation}
where $\tilde{t}_{\hat{x}+\hat{y}}=\tilde{t}_{\hat{x}+\hat{y}}=\tilde{t}_2$.

We define $\hat{D}^\dag_{\bf m,\hat{\bf y}}=\hat{b}^\dag_{\bf m}\hat{b}_{\bf m-\hat{y}}$, the creation operator of a dipole along the $y$ direction. The kinetic energy of the dipole can be written as
\begin{eqnarray}\label{dipolekinetic}
\begin{aligned}
  \hat{\tilde{K}}_D&=-\sum_{\bf m}\big\{t'_2\hat{D}^\dag_{{\bf m+\hat{x}},\hat{\bf y}}\hat{D}_{{\bf m},\hat{\bf y}}+t_2\hat{D}^\dag_{{\bf m+\hat{y}},\hat{\bf y}}\hat{D}_{{\bf m},\hat{\bf y}} \\
  &+\tilde{t}_2(\hat{D}^\dag_{{\bf m+2\hat{x}},\hat{\bf y}}+\hat{D}^\dag_{{\bf m+\hat{x}+\hat{y}},\hat{\bf y}}+\hat{D}^\dag_{{\bf m+\hat{x}-\hat{y}},\hat{\bf y}})\hat{D}_{{\bf m},\hat{\bf y}}+h.c.\big\}
  \end{aligned}
\end{eqnarray}

We note that, in the limit where $U$ is much larger than any other energy scales, $t'_{2}\rightarrow \frac{2V}{V^2-\Delta^2_x}t_1^2$, $t_2\rightarrow \frac{1}{V+\Delta_y}t_1^2$, and $\tilde{t}_2\rightarrow 0$. In this limit, only the nearest neighbor tunnelings exist but a dipole can tunnel in both directions, being a superposition of a lineon and a planon.  

To create a tensor gauge field, similar to discussions in the main text, an extra small quadratic potential in the $x-y$ plane can be added, $\hat{H}'_B=\hat{H}_B+\hat{V}_L$ where
\begin{equation}\label{quad_potential}
  \hat{V}'_B=\frac{1}{2}\sum_{\bf m}(\alpha m_x+\beta m_y)^2\Delta'\hat{n}_{\bf m},
\end{equation}
where $\alpha$ and $\beta$ are both dimensionless coefficient to control the direction of quadratic potential which satisfying $\alpha^2+\beta^2=1$. Using a unitary transformation to eliminate $\hat{V}_B+\hat{V}'_B$, the single-particle tunneling acquires an additional time-dependent phase,
\begin{eqnarray}\label{unit_trans_kinetic}
\begin{aligned}
  \hat{\tilde{K}}_B=&-\sum_{\bf m}(t_1e^{-i\Delta_xt-i(m_x+1/2)\alpha^2\Delta't-im_y\alpha\beta\Delta't}\hat{b}^\dag_{\bf m}\hat{b}_{\bf m+\hat{x}} \\
  &+t_1e^{-i\Delta_yt-im_x\alpha\beta\Delta't-i(m_y+1/2)\beta^2\Delta't}\hat{b}^\dag_{\bf m}\hat{b}_{\bf m+\hat{y}}+h.c.)
  \end{aligned}
\end{eqnarray} 
In the limit $\Delta'\ll\Delta$, the Floquet theory gives rise to a ring-exchange interaction, 
\begin{equation}
  \hat{\tilde{K}}_{P}=-\sum_{\bf m}(t_2'e^{-i\alpha\beta\Delta't}\hat{b}^\dagger_{\bf m}\hat{b}^\dagger_{{\bf m}+\hat{\bf x}+\hat{\bf y}}\hat{b}_{{\bf m}+\hat{\bf x}}\hat{b}_{{\bf m}+\hat{\bf y}}+h.c.).
\end{equation} 
Similar discussions apply to other tunnelings.

\end{document}